\documentclass[aps,twocolumn, prl,superscriptaddress, longbibliography, notitlepage]{revtex4-2}
\usepackage{bm,amsmath,graphicx, mathtools,multirow,gensymb}
\usepackage[colorlinks=true,linkcolor=MidnightBlue,urlcolor=black,citecolor=MidnightBlue,anchorcolor=MidnightBlue]{hyperref}
\usepackage[dvipsnames]{xcolor}
\begin{document}
\title{Active phase separation: role of attractive interactions from stalled particles}
\author{Kingshuk Panja}
\thanks{ph20d205@smail.iitm.ac.in}
\affiliation{Department of Physics, Indian Institute of Technology Madras, Chennai, India}
\author{Rajesh Singh}
\thanks{rsingh@physics.iitm.ac.in}
\affiliation{Department of Physics, Indian Institute of Technology Madras, Chennai, India}
\begin{abstract}
Dry active matter systems are well-known to exhibit Motility-Induced Phase Separation (MIPS). However, in wet active systems, attractive hydrodynamic interactions mediated by active particles stalled at a boundary can introduce complementary mechanisms for aggregation. In the work of Caciagli \emph{et al.} (PRL 125, 068001, 2020), it was shown that the attractive hydrodynamic interactions due to active particles stalled at a boundary can be described in terms of an effective potential. 
In this paper, we present a model of active Brownian particles, where a fraction of active particles are stalled, and thus, mediate inter-particle interactions through the effective potential.
Our investigation of the model reveals that a small fraction of stalled particles in the system allows for the formation of dynamical clusters at significantly lower densities than predicted by standard MIPS. We provide a comprehensive phase diagram 
in terms of weighted average cluster sizes that is mapped in the plane of the fraction of stalled particles ($\alpha$) and the Péclet number. Our findings demonstrate that even a marginal value of $\alpha$ is sufficient to drive phase separation at low global densities, bridging the gap between theoretical models and experimental observations of dilute active systems.
\end{abstract}
\maketitle

The study of matter traditionally focuses on systems at thermodynamic equilibrium, where structures and phases are governed by minimizing the free energy \cite{chaikin2000principles}. However, the last couple of decades have seen the emergence of active matter as a distinct field, comprising constituents that consume energy locally to generate persistent motion, driving the system far from equilibrium \cite{ramaswamy2017active, marchetti2013, cates2025active, van2024soft}. Examples span from living systems, such as bacterial colonies \cite{petroff2015fast}, and motile cells \cite{tan2022odd}, to artificial systems like self-propelled droplets \cite{kumar2024emergent} and catalytic Janus particles \cite{ebbens2010pursuit}. 
Understanding the collective behaviors that emerge from simple  
rules in active matter systems is a major research goal of
modern non-equilibrium statistical physics. Apart from academic interest, 
the knowledge base acquired by studying these active systems
can inform the design of novel functional materials
whose properties can be tuned. 
\\

A remarkable collective phenomena observed in
dry active matter systems is Motility-Induced Phase Separation (MIPS) \cite{cates2015, marchetti2016minimal, digregorio2018full,chacon2022, sese2022}. In MIPS, particles spontaneously separate into dense and dilute phases, despite the lack of any attractive inter-particle interactions. MIPS is driven purely by the self-propulsion or the motility of the particles.
%
Usually, particle-based models studying MIPS ignore role of hydrodynamic interactions between particles. 
The role of hydrodynamic interactions and effect of boundaries in the fluid for the
phase separation of active particles can be understood from the complimentary mechanism, where the 
attractive flow induced by particles swimming into a boundary leads to aggregation \cite{singh2016crystallization, drescher2009, petroff2015fast, tan2022odd, caciagli2020controlled, aubret2018targeted, squiresBrenner2000}.
The main operational principle:
active particles swimming into a fluid boundary (such as wall), and thus stalled, create attractive fluid flow that causes flow-induced phase separation of active particles \cite{singh2016crystallization}. 
In \cite{caciagli2020controlled}, it was shown that the attractive flow-mediated inter-particle interactions 
can be written as  an effective attractive potential, which depends on self-propulsion speed of the particle and viscosity contrast of the interface at which phase separation occurs. 
To the best of our knowledge, a model exploring the role of this effective potential in the phenomenology of MIPS has not been studied. \\

\begin{figure}[b]
    \centering
    \includegraphics[width=0.475\textwidth]{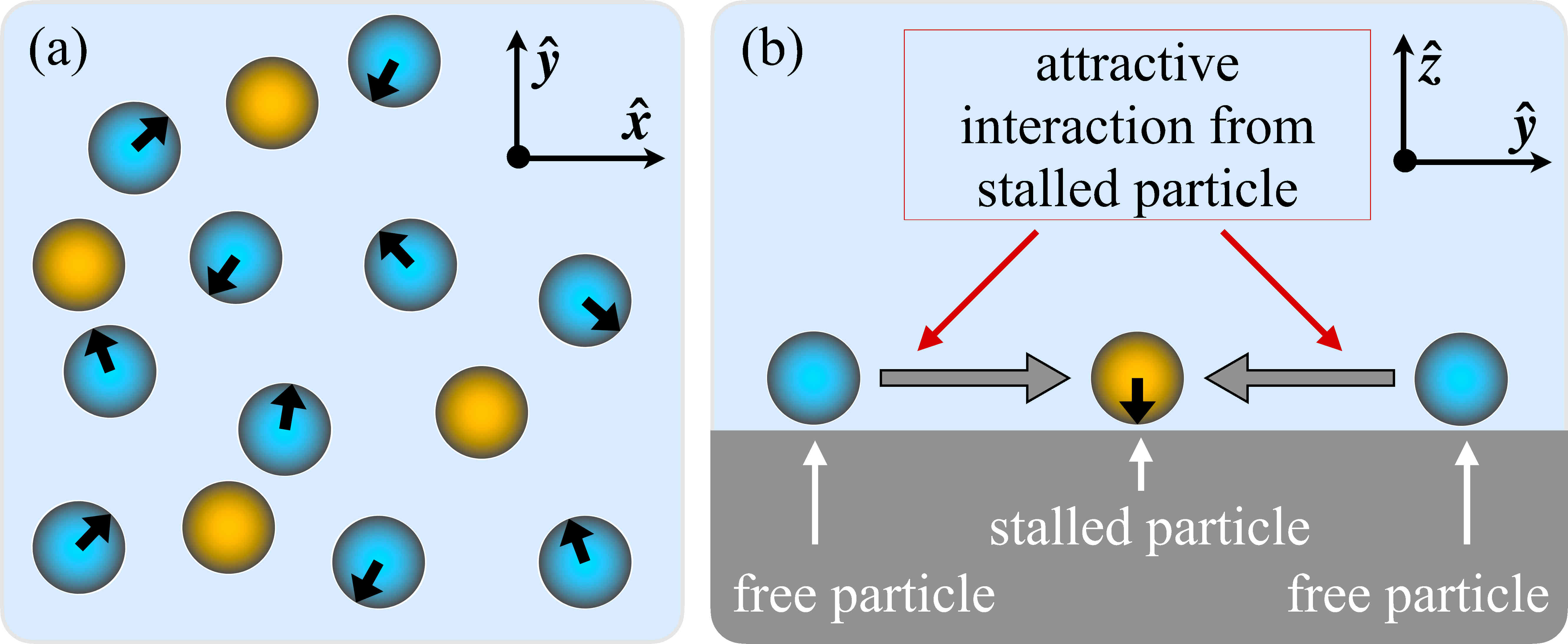}
    \caption{Panel (a) shows a schematic diagram of motion of free (blue) and stalled particles (yellow) in the $xy$-plane. The stalled particles have no self-propulsion in the plane while free particles can freely self-propel. (b) shows that the stalled particles acts as sites of attraction in the system. 
    See Eq.\eqref{eq:HI_pot} for the functional form of effective inter-particle attractions mediated by stalled particles.
    }
    \label{fig:schematic}
\end{figure}

In this paper, we study a system of active Brownian particles in two-dimensions, where we investigate the influence of an effective attractive interaction due to a small fraction $\alpha$ of particles that are stalled (see Fig.\eqref{fig:schematic}) at a boundary with viscosity contrast $\lambda$. Using extensive numerical simulations of the model, we show that the phase separation of active particles is drastically modified 
even at marginal values of $\alpha$. In addition, we find the 
morphology of cluster to depend on $\alpha$, apart from Peclet number and density.
The clusters obtained from stalled particles are more stable than those obtained by MIPS.
Overall, we find that the increasing $\alpha$ makes system more crystalline.
We also find more pronounced clustering for $\lambda=0$ than at $\lambda\rightarrow\infty$ for the same value of $\alpha$, signifying the role of boundary conditions in phase separation. Our results indicate that the phase separation can proceed  at significantly lower densities and activity 
than predicted by standard MIPS on addition of small fraction of stalled particles. We describe our model and results below. \\

%

\textit{Model:}
We consider a system of $N$ active colloidal particles of radius $b$
with intrinsic speed $v_{\scriptscriptstyle{0}}$. 
A schematic diagram of our system is present in Fig.\eqref{fig:schematic}. 
We consider $N_f$ free active particles, which can freely self-propel in two-dimension on a surface.  
The position of the $i^\text{th}$ free particles is denoted by $\mathbf r^\mathrm{f}_{i}$, while its orientation is denoted by 
$
\mathbf e^{\mathrm{f}}_i = \cos\theta^\mathrm{f}\,\hat x + \sin\theta^\mathrm{f}\,\hat y$.
 We also consider $N_s$ stalled particles, such that $N_f+N_s=N$.
The position  of the $i^\text{th}$ stalled particle is denoted as: $\mathbf r^\mathrm{s}_{i}$.
The stalled particle only have positional dynamics, while the free particles have dynamics of both positions and orientations in the two-dimensional space.
The update equations for free and stalled particles:
\begin{align}
&\dot{\mathbf r}^\mathrm{f}_{i} =v_{\scriptscriptstyle{0}}\mathbf e^\mathrm{f}_i+\mu_t\mathbf F^\mathrm{f}_i,\quad 
     \dot{\theta^\mathrm{f}_{i}}  = \sqrt{2D_{r}}\,\xi_r,
\qquad 
\dot{\mathbf r}^{\mathrm{s}}_{i}=\mu_t\,\mathbf F^{\mathrm{s}}_i.
\label{eq:posDyn}
\end{align}
 Here $v_{\scriptscriptstyle{0}}$ is the self-propulsion speed of the 
 free particles, $\mu_t=1/(6\pi\eta_1 b)$ is the mobility, while $D_r$ is the rotational diffusion constant and $\xi_r$ is a
 stochastic variable with zero mean and unit variance.\\
 
In Eq.\eqref{eq:posDyn}, $\mathbf F_i^{\mathrm{f}}$ and $\mathbf F_i^{\mathrm{s}}$ are respectively, the inter-particle forces on the $i$th free and stalled particles, that can be written as: 
 $\mathbf F^{\mathrm{f}}_i = \sum_{j = 1}^{N_s}\mathbf F^\text{HI}_{ij} + 
    \sum_{
    \substack{j = 1,\,j\neq i}
    }^{N_f+N_s} \mathbf F^\text{WCA}_{ij}$ is the force on free particles due to free and stalled particles, while we have the following expression
    for stalled particles: 
    $   \mathbf F^{\mathrm{s}}_i = \sum_{
    \substack{j = 1,\, j\neq i}
    }^{N_s}\mathbf F^\text{HI}_{ij} + 
    \sum_{
    \substack{j = 1,\,j\neq i}
    }^{N_f+N_s} 
    \mathbf F^\text{WCA}_{ij}
$.
Here, $\mathbf F^\text{WCA}_{ij} = -\mathbf \nabla_i \Phi^\text{WCA}(\mathbf r_i - \mathbf r_j)$ to ensure that particles do not overlap with each other 
\cite{weeks1971role}. It is defined as: 
$ \Phi^{\mathrm{WCA}}(r)= 
    \epsilon\left(\frac{r_\text{m}}{r}\right)^{12} - 2\epsilon\left(\frac{r_\text{m}}{r}\right)^6 + \epsilon$
    {when $r\leq r_\text{m}$, while it vanishes otherwise.
The attractive hydrodynamic interactions due to stalled particles can be written in terms of an effective potential such that the force is:
$\mathbf F^\mathrm{HI}_{ij} = -\mathbf\nabla_i \Phi^\mathrm{HI}(\mathbf r_i , \mathbf r^\mathrm{s}_j)$, where the effective potential $\Phi^\mathrm{HI}$  
is given as \cite{caciagli2020controlled}:
\begin{align}
\Phi^{\mathrm{HI}}(\mathbf r_i , \mathbf r^\mathrm{s}_j
) 
= -\frac{v_{\scriptscriptstyle{0}}}{4\pi\eta_1\mu_\perp\mu_\parallel}\left[\frac{1}{1+\lambda}\frac{1}{d } + \frac{2\lambda}{1+\lambda}\frac{1}{{d}^3}\right]
\label{eq:HI_pot}
\end{align}
Here $\lambda=\eta_2/\eta_1$ is is the ratio of the viscosity of the two fluids while the particles are confined at the interface, $h$ is the height of the stalled particles (in our model, we also have put free particle at the same height). 
Here, $d= \sqrt{|\mathbf r_i -\mathbf r^s_j|^2 + 4h^2|}/h$. In this paper, we have taken $h = 2b$. The particles are in a fluid of viscosity $\eta_1$, while the medium on other side of the boundary has viscosity $\eta_2$. Thus, for a fluid-solid boundary, such as a solid wall, $\lambda\rightarrow\infty$, while $\lambda=0$ for an air-water 
interface.  
Here $\mu_\parallel$ ($\mu_\perp$) is the mobility for the motion of a sphere parallel (perpendicular) to the boundary. See SI \cite{siText} for their explicit forms. Note that the strength of this attractive potential is determined by self-propulsion speed $v_{\scriptscriptstyle{0}}$, since it originates from the flow field produced by a stalled particle.
The value of $\lambda$ corresponds the fluid-solid boundary, it is $\lambda\rightarrow\infty$ for plane wall bounding the fluid with no-slip boundary condition. In this case, the attractive interaction due to the potential of Eq.\eqref{eq:HI_pot} decays as $1/R^3$, where $R$ is the distance between the colloids. Thus, the attractive inter-particle forces decay as $1/R^4$. On the other hand, for the case of $\lambda=0$, the potential decays as $1/R$, while the attractive inter-particle forces decay as $1/R^2$. To summarize, our model has two kind of particles: free and stalled active particles. The free particles have both positional and orientational dynamics, while the orientations of the stalled particle is fixed: it point into the two-dimensional substrate, on which particles are located. This induces a net attractive interactions, through the effective potential of Eq.\eqref{eq:HI_pot}, between particles.  
\\ 

\begin{figure*}[t]
    \includegraphics[width=0.94\textwidth]{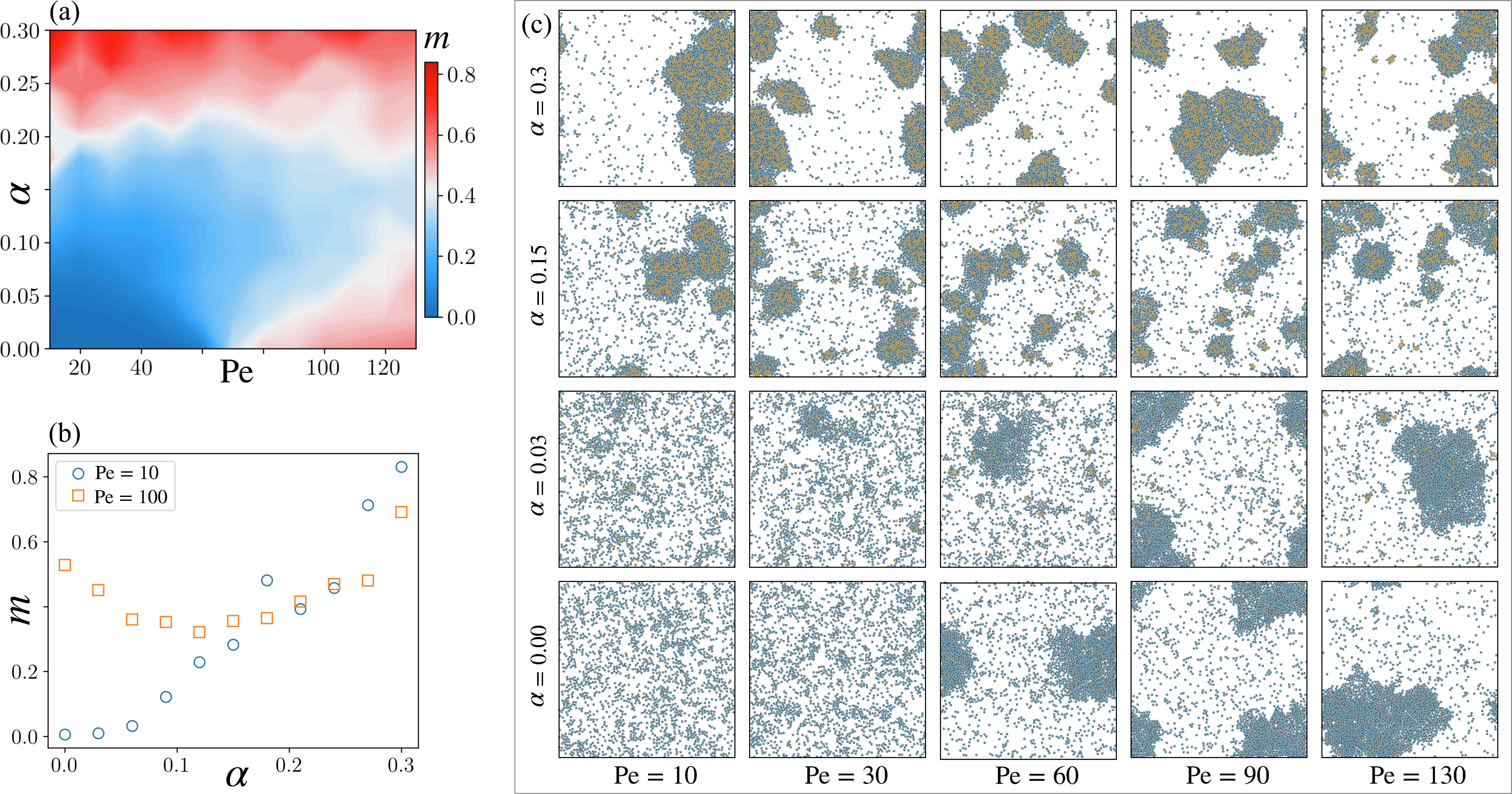}
\caption{Active phase separation with a fraction of stalled particles ($\alpha$) for the case of $\lambda\rightarrow\infty$ and area fraction 
$\phi=0.3$.
(a) Phase diagram in terms of
weighted average cluster size $m$  in the plane of $\alpha$ and Péclet number Pe.
(b) Plot of $m$ as a function of $\alpha$. We obtain clustering at Pe=$10$ as $\alpha$ is increased, a region where clustering is not possible from usual MIPS. 
Interestingly, at Pe=$100$, cluster size first decreases on increasing $\alpha$. Thus, suppressing MIPS due to reduction in number of particles self-propelling in the plane. 
But on further increasing $\alpha$, we 
again get pronounced aggregation due to dominance of attractions mediated by stalled particles. 
(c)
    Snapshots from steady state at different values of Péclet number Pe and $\alpha$. 
    It should be noted that we get phase separation at any value of Pe and $\phi$ on increasing $\alpha$ beyond a threshold.
    \label{fig:Lambda_INFTY}
}
\end{figure*}
\begin{figure}[b]
    \centering
    \includegraphics[width=0.98\linewidth]{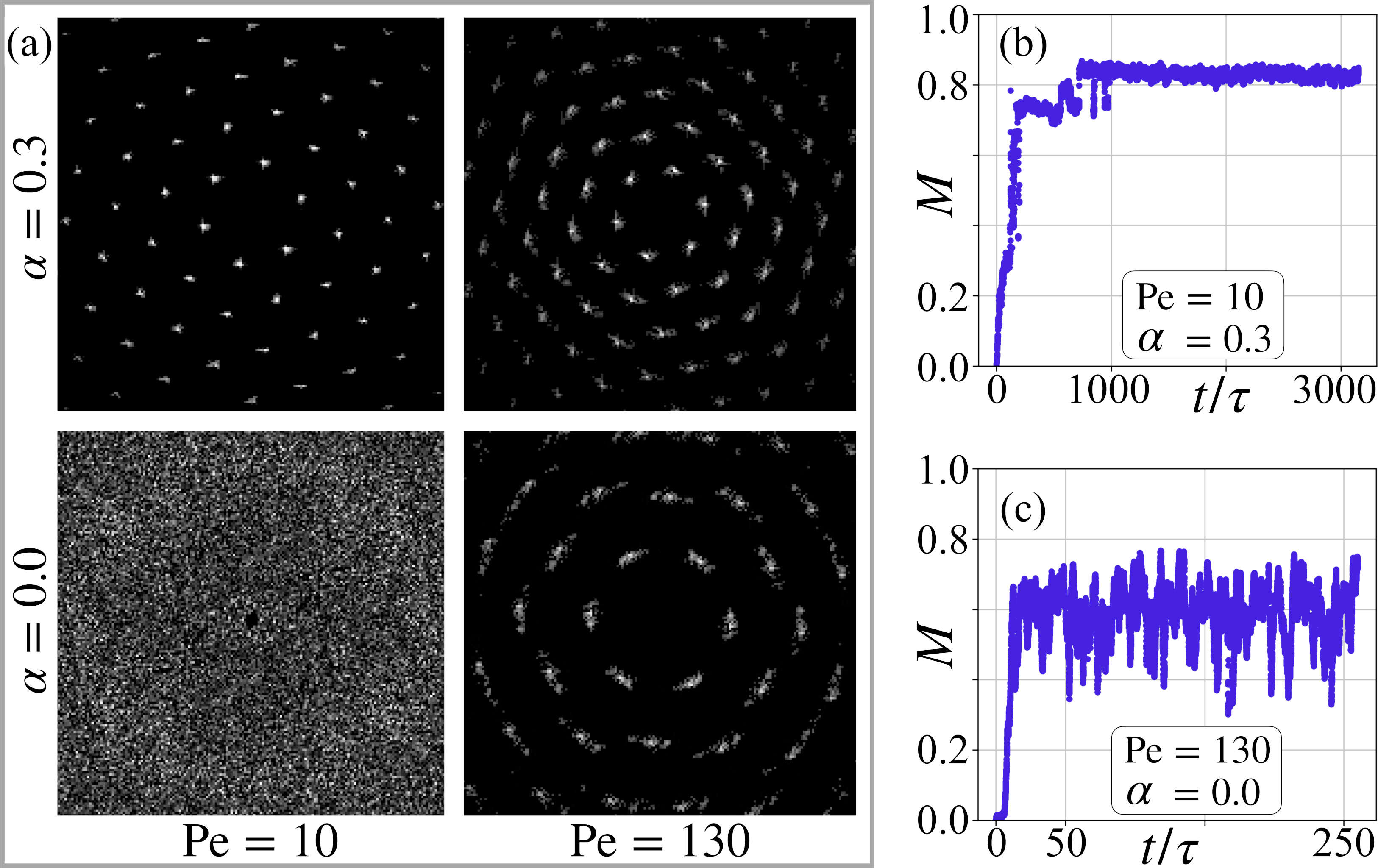}
    \caption{Panel (a)  contains 
    plots of the structure factor $S(\mathbf k)$ at corresponding values of Pe and $\alpha$, while 
    the contribution from $\mathbf k=0$ is discarded.
    There is no clustering at $\alpha=0$ and Pe = 10.
    (b-c) shows the growth of weighted average cluster size $M$. (b) shows growth solely 
    due to effective attraction from potential of Eq.\eqref{eq:HI_pot} as
    $\alpha=0.3$ and Pe=10. 
(c) shows growth through MIPS alone as $\alpha=0$ and Pe=130. }
    \label{fig:M_evolution}
\end{figure}
%

To study the system, we define key dimensionless parameter using equations of motion.
The Péclet number Pe and ratio of stalled particle number and total number of particles $\alpha$. They are:
\begin{align}
    \mathrm{Pe} = v_{\scriptscriptstyle{0}} \frac{\tau}{b},\qquad 
     \alpha = \frac{N_s}{N_s + N_f} = \frac{N_s}{N}
\end{align}
Here, $\tau = {b^2}/{D_r}$ is the persistence time.
In this paper, we have measured weighted mean cluster size by $\langle n\rangle_w = \sum_n n\,p_w$, where $n$ is the number of particles in the cluster and $p_w$ is the weighted cluster size distribution, defined as $p_w(n) = n\,N_n/(\sum_n nN_n)$. Here, $N_n$ is the number of clusters of size $n$. 
 We map the phase diagram using an order parameter $m$, which is the weighted average cluster size that is also averaged over the steady-state $m$. To this end, we use the average number of particles $M$ in cluster. These are defined as:
 \begin{align}
     m  =\langle M \rangle_{\mathrm{ss}},\qquad M=\frac{\langle n \rangle}{N}.
 \end{align}
Here $N$ is total number of particles and $\langle\dots \rangle_{\mathrm{ss}}$ implies that the average is also taken over the steady-state. \\

\textit{Phase separation for $\lambda\rightarrow\infty$:
}
We first describe our results for motion of 
active colloids in a two-dimensional plane at 
$\lambda\rightarrow\infty$ (which corresponds to a fluid-solid boundary, such as a plane wall). Our results for this case are summarized in Fig.\eqref{fig:Lambda_INFTY}.
The interplay between MIPS and effective attractions due to stalled particles creates a complex phase diagram.
The phase diagram is given in Fig(\ref{fig:Lambda_INFTY})a. It can be clearly seen that the phase separation happens at all values of Péclet number once the fraction of stalled particles is increased beyond a critical number. Interestingly, MIPS is initially suppressed on increasing the fraction of stalled particles, as shown in Fig(\ref{fig:Lambda_INFTY})b. This is because the effective number of free particles is reduced. But, on increasing the number of stalled particles, we again obtain full clustering. Snapshots from steady-state are shown in Fig(\ref{fig:Lambda_INFTY})c. 
The kinetics of phase separation
obtained from numerical solutions is shown in movie 1 
and SI Fig 1. We show that the
uniform state is destabilized, on increasing the fraction of the stalled particles $\alpha$, at 
any value of Péclet number.
On increasing $\alpha$, the attractive interaction 
between the particles is balanced by
steric inter-particle repulsion between 
to produce 
crystallites with hexagonal positional order. Indeed, we find that the hexagonal order in the system is enhanced on increasing $\alpha$. This is shown in Fig.\eqref{fig:M_evolution}a. 
It should be noted for cases where phase separation occurs, the structure factor $S(\mathbf k)$ is computed from interior of clusters. 
In Fig(\ref{fig:M_evolution})b and c, we show that the evolution of cluster size $M$ is more stable - at a given Peclet number - when we are increasing the fraction of stalled particles. 
\\

In purely repulsive active systems, MIPS occurs because self-propelled particles accumulate in regions where their local velocity is reduced by collisions, leading to a feedback loop of slowing and packing \cite{cates2015}. In our 
model with a small fraction of stalled particles, the attractions to stalled particles
act as a ``glue" that stabilizes the aggregates.
While MIPS alone requires high Péclet numbers for phase separation, 
the addition of attractive forces can facilitate the formation of stable clusters even at lower Peclet numbers. 
However, the morphology of these clusters can differ significantly: whereas pure MIPS typically produces clusters that are maintained by a constant flux of active particles, the presence of attractions leads to the formation of rigid, kinetically trapped aggregates or crystalline structures that resist the fluidizing effect of high activity. See Fig.\eqref{fig:M_evolution}, where we show structure factor and time evolution of cluster size. 
In this paper, we have assumed that the stalled particles continue to point into the boundary and thus act as persistent sites of attraction. It is possible that they turn away during the course of an experiment, and thus, can lead to break-up of clusters. Such a route provides a mechanism for local fragmentation of clusters.
\\

\begin{figure}[t]
    \includegraphics[width=0.98\linewidth]{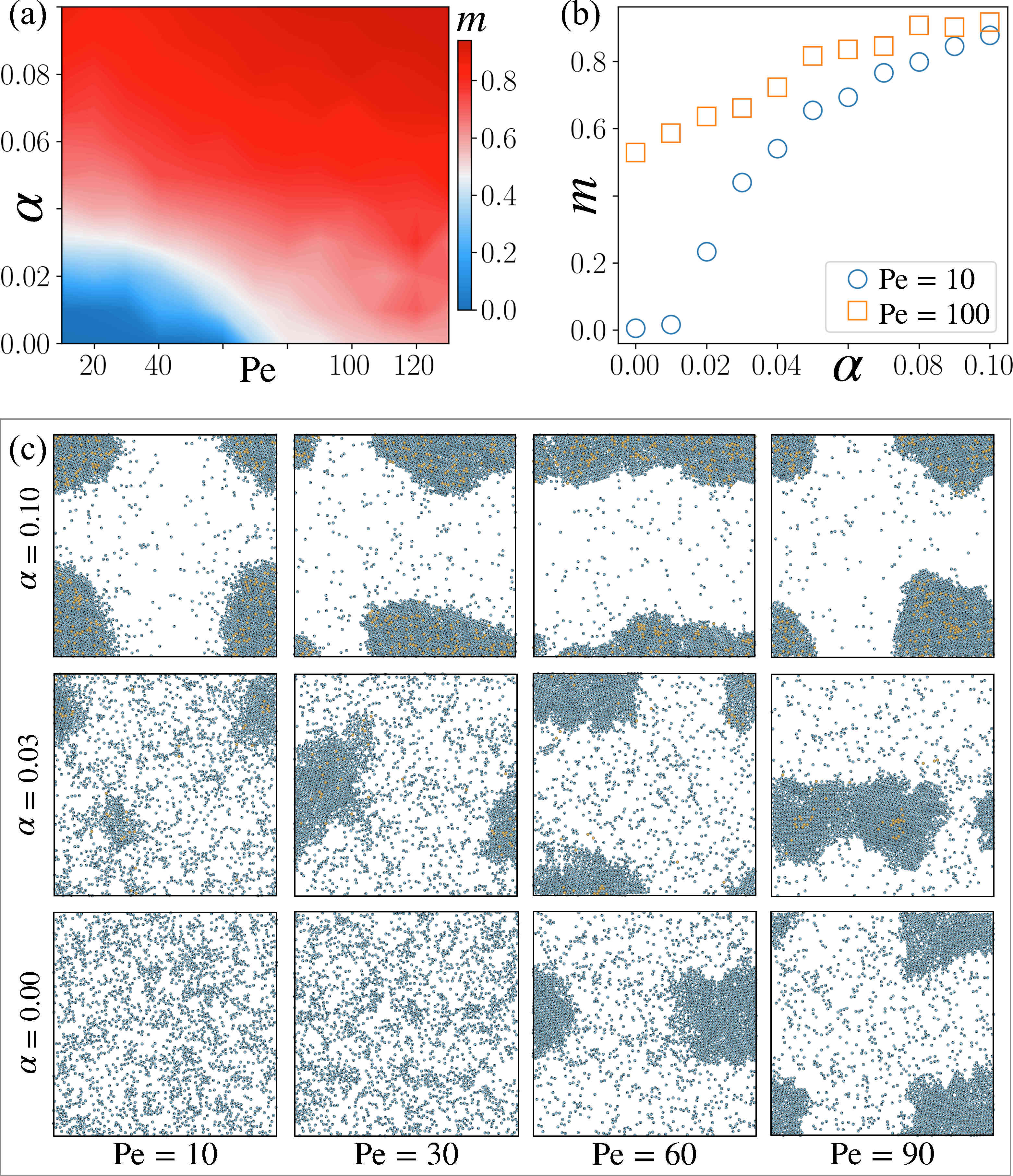}
\caption{Active phase separation with a fraction ($\alpha$) of stalled particles for the case of $\lambda=0$. Here, the total area fraction is $\phi=0.3$.
(a) Phase diagram in terms of weighted average cluster size $m$ 
in the plane of $\alpha$ and Péclet number Pe. 
We get phase separation at any value of Pe and $\phi$ on increasing $\alpha$ beyond a threshold. 
(b) Plot of $m$ for two different Pe as a function of fraction of stalled particles. (c)
    Snapshots from steady state at different values of Péclet number Pe and $\alpha$. 
\label{fig:Lambda_ZERO}
}
\end{figure}

\textit{Phase separation for $\lambda=0$:}
\label{sec:lambdaZero}
For the motion of active particles in a two-dimensional surface, such as a air-water interface, we have $\lambda=0$.  
In this case,
the effective potential of Eq.\eqref{eq:HI_pot} decays as $1/R$. Thus, effective attraction due to stalled particles is more long-ranged compared to the case of $\lambda\rightarrow\infty$. 
The phase diagram and snapshots from steady-state for this case in given in 
Fig.\eqref{fig:Lambda_ZERO}a and Fig.\eqref{fig:Lambda_ZERO}c respectively. It can be seen that even with very small fraction of particles - up to $3\%$ we obtain full phase separation due to long-ranged nature of the attraction. This is clearly 
seen in the plot of average cluster size $m$ in Fig.\eqref{fig:Lambda_ZERO}b. 
Here, we obtain clustering at $\phi=0.3$ and $\mathrm{Pe}=10$,
as $\alpha$ is increased, 
which is not possible from MIPS alone.
It is worthwhile to 
compare 
the results of Fig.\eqref{fig:Lambda_ZERO} - enumerating our finding for $\lambda=0$ -
with the phase diagram of Fig.\eqref{fig:Lambda_INFTY} where $\lambda\rightarrow\infty$. 
For the case of $\lambda=0$, we need a smaller fraction of stalled particles for phase separation as the interactions decay more slowly $(\Phi^{\mathrm{HI}}\sim 1/R)$, 
while in the case of $\lambda\rightarrow\infty$, we have: $\Phi^{\mathrm{HI}}\sim 1/R^3$. Consequently, for a given set of parameters, the clustering tendency and cluster size is higher for $\lambda=0$, which corresponds to an air-water interface \cite{thutupalli2018FIPS}. 
At intermediate values of $\lambda$, such as at an oil-water interface \cite{caciagli2020controlled}, we expect the phase separation dynamics to be 
between these two extreme cases of $\lambda\rightarrow\infty$ and $\lambda=0$. 
We leave a detailed analysis of the intermediate values of $\lambda$ for a future work.  \\


\textit{Conclusion:} Phase separation in dry active matter systems is through MIPS, while attractive flow can drive phase separation in wet active matter systems. This paper studies the subtle interplay of these two mechanism using a model 
where a fraction of stalled particles mediate attractive interactions through an effective hydrodynamic potential.   
We show that phase separation is possible even at smaller value of area fraction and activity, than those from standard MIPS. Moreover, we show that MIPS is first suppressed on addition of stalled particles (due to reduction of free particles), but on further increasing the number of stalled particles,
inter-particle
attractions leads to aggregation of particles. Moreover, crystalline order in the system is enhanced
 on increasing the number of stalled particles.  
It would be interesting to verify our results and phase diagrams in experimental systems where we can control the number of stalled particles. 
Such a control on achieving stalled particles in an active suspension 
has been recently demonstrated
in experimental systems of active particles 
\cite{aubret2018targeted, martinetPRX2025}. \\

Our results complement other theoretical studies on phase separation of active particles using particle-based model of dry active matter \cite{redner2013reentrant, redner2013structure, stenhammar2014phase, garcia2025dynamics, solon2018, demin2017, tung2016micro} and models where explicit hydrodynamic interactions has been included \cite{singh2016crystallization, thutupalli2018FIPS, zhou2025hydrodynamic, matas2014hydrodynamic}.  This work advances understanding of non-equilibrium active matter by linking MIPS and effective model of hydrodynamic interactions.  
Our findings may also be relevant to studies in which a continuum modeling is employed to study phase separation of active particles \cite{stenhammar2013PRL, singh2019self, cates2025active}. \\

\emph{Acknowledgments:}
 We thank R Adhikari, ME Cates, E Eiser, PBS Kumar, I Pagonabarraga, S Thutupalli, and C Valeriani for helpful discussions.
%

\end{document}